\documentclass[trans]{IEEEtran}
\usepackage[table,xcdraw]{xcolor}
\usepackage{hyperref}
\usepackage{lettrine}
\usepackage{epsfig}
\usepackage{float}
\usepackage{footnote}
\usepackage{cite}
\usepackage{float}
\usepackage{soul}
\usepackage{amssymb}
\usepackage{amsthm}
\usepackage{amsmath,mathabx}
\usepackage{color}
\usepackage{epstopdf}
\usepackage{tabularx,tabulary}
\usepackage[font=footnotesize]{caption}
\usepackage{graphicx}
\usepackage{caption}
\usepackage{subcaption}
\usepackage{fontawesome}
\usepackage{cuted}
\usepackage{dblfloatfix}    
\usepackage{bbm}
\usepackage{algorithm,algorithmic}
\usepackage{wrapfig}
\usepackage{tcolorbox}
\usepackage[normalem]{ulem}
\usepackage{multirow}
\usepackage{graphicx}

\graphicspath{{../}}

\definecolor{maroon}{RGB}{186,0,0}
\definecolor{pembe}{RGB}{96,26,149}
\definecolor{mavi}{RGB}{115,212,248}
\definecolor{haki}{RGB}{38,99,33}

\begin{document}
\title{Topology Optimization for 6G Networks: \\ A Network Information-Theoretic Approach }
\author{
Abdulkadir~Celik,~\IEEEmembership{Senior Member,~IEEE,} Anas~Chaaban,~\IEEEmembership{Senior Member,~IEEE,}\\
Basem~Shihada,~\IEEEmembership{Senior Member,~IEEE,} and~Mohamed-Slim~Alouini,~\IEEEmembership{Fellow,~IEEE}
\thanks{A.Celik, B. Shihada, and M.-S. Alouini  with  Computer,  Electrical,  and  Mathematical  Sciences and Engineering (CEMSE) division at King Abdullah University of Science and Technology (KAUST),  Thuwal,  23955-6900, KSA. A. Chaaban is with School of engineering at University of British Columbia (UBC).}
}
\maketitle
\thispagestyle{empty} 
\pagestyle{empty} 
\begin{abstract}
The classical approach of avoiding or ignoring interference in wireless networks cannot accommodate the ambitious quality-of-service demands of ultra-dense cellular networks (CNs). However, recent ground-breaking information-theoretic advances changed our perception of interference from a foe to a friend. This paper aims to shed light on harnessing the benefits of integrating modern interference management (IM) schemes into future CNs. To this end, we envision a hybrid multiple access (HMA) scheme that decomposes the network into sub-topologies of potential IM schemes for more efficient utilization of network resources. Preliminary results show that HMA can multiply non-orthogonal multiple access performance, especially under dense user deployment.
\end{abstract}
\section*{ \centering\textbf{Introduction}}
\label{sec:intro}

\lettrine{I}{n} a recent study, CISCO predicted that smart-phones' data-traffic will surpass that of personal computers by 2022 \cite{cisco}, effectively tripling between 2017 and 2022. It also anticipated a high growth in traffic generated by tablets and machine-type communications. Altogether, wireless devices will constitute close to two-thirds of the total traffic by 2022. While this might not be surprising to some, the alarming part is that data communication capabilities will only nearly double by 2022 compared to 2017! Since our daily activities continue to become increasingly dependent on data communications, one can only imagine the consequences of wireless networks' inability to cope with data-demand on our lives. As a result, 6G networks are expected to further improve the 5G goals of enhanced mobile broadband, massive machine-type communication, and ultra-reliable low-latency communication. However, these goals turned into being extremely challenging as we approached to the fundamental limits of classical communications paradigms.

Even though it is believed that cell-densification can solve this problem, deploying more cells per Km$^2$ yields an interference-limited network. Classical schemes that avoid interference by allocating users on non-interfering channels become helpless in accommodating the ambitious demands of future networks. First, dividing the limited network resources over a large number of users decomposes the cellular network (CN) topology into one consisting of a set of point-to-point (P2P) channels with small bandwidth per channel. Second, inter-cell interference due to frequency-reuse becomes devastating given the network density. Instead, more emphasis should be on schemes that can contain interference, thus allowing beyond P2P sub-topologies. Along this line, researchers recently started to allow more tolerance for interference using NOMA, which decomposes the CN into multiple access channel (MAC) and broadcast channel (BC) sub-topologies in the uplink (UL) and downlink (DL), respectively. Despite its improvement upon classical orthogonal multiple-access (OMA), this transition remains conservative in terms of the sub-topologies it allows in a CN. Moreover, this transition relies on 4-decade-old results from information theory (IT) related to the MAC and BC \cite{ElgamalKim}. 

We strongly believe that a more liberal and updated approach can lead to substantial improvements. Developments in network IT over the past two decades highlighted interference management (IM) schemes that led to ground-breaking results in IT and fundamentally changed our perception of interference. These schemes transform interference from a foe to a friend and theoretically increase the data rate significantly. The main question remains on the practicality of these schemes, the gain they can provide under real-life constraints, and how sub-topologies favoring these schemes can be orchestrated in dense and heterogeneous CNs. In this paper, we accordingly aim to shed light on the integration of recent network IT advances into CN optimization to harness gains of modern IM schemes. To this end, we will present our vision towards diversifying and hybridizing potential network sub-topologies for a more efficient utilization of CN resources.

In the remainder of this paper, we first give background and motivation for embracing new IM schemes. Then, we provide the proposed network IT approach as well as potential IM schemes to be used in future networks. Thereafter, a hybrid multiple access (HMA) concept is presented along with results for proof of concept, which is followed by a discussion of potential mathematical tools for network topology optimization. Finally, we conclude the paper with a few remarks.

\section*{ \centering\textbf{IT Background and Motivation}}
\label{sec:motivation}
Classical CN resource allocation aims at \textit{`interference avoidance'} by using OMA to isolate communicating pairs by dedicating time/frequency resource blocks (RBs) to a user-equipment (UE) base-station (BS) pair in the UL or DL. Thus, the CN is decomposed into a set of non-interfering P2P channels. Based on P2P channel theory, the data rate per RB in OMA can be expressed by  
\begin{equation}
\label{eq:rate}
\rm{R}=\frac{\rm{W}}{\rm{N}}\times \log_2 \left( 1+ \frac{\rm{S}}{\rm{I}+\sigma}\right) \: [\mathrm{bps}]
\end{equation}
where $\rm{W}$ is the total bandwidth, $\rm{N}$ is the number of RBs in $\rm{W}$, $\rm{S}$ is the signal power, $\rm{I}$ is the interference power from transmitters using the same RB, and $\rm{\sigma}$ is the noise power within the RB. 

To meet the ever-increasing data demand, one has to increase $\rm{R}$, that amounts to increasing the spectral efficiency (SE), i.e., $ \log_2 \left( 1+ \frac{\rm{S}}{\rm{I}+\sigma}\right)$, and/or decreasing the number of RBs, $\rm{N}$. In the past, researchers focused on OMA and developed codes and resource allocation schemes to maximize the first objective. However, connecting a massive number of devices using OMA decreases $\rm{R}$ since this requires $\rm{N}$ to be huge. So the following question arises: How can we decrease $\rm{N}$ while connecting a large number of devices?\footnote{We will use UE henceforth to refer to a generic connected device.}

A practical recommendation is to deploy more small-cells over a macro-cell coverage area to decrease the required $\rm{N}$ per cell. Cell densification enables allocating more RBs per UE for a larger communication bandwidth, but the expense is an increased intra-cell interference due to the dense spatial reuse. 
Another recommendation is to allow a cluster of UEs to use the same RB within a cell, which decreases the required $\rm{N}$ on the one hand yet introduces inter-cell interference on the other hand. 

Thus, both recommendations have a side effect; they increase $\rm{I}$ and hence decrease the SE. Seemingly, there is a conflict between the first and the second objectives since decreasing the number of RBs necessitates UE resource sharing and hence increases interference. However, this is only problematic if under the classical approach of ignoring interference, also known as \textit{`Treating Interference as Noise'} (TIN). Using TIN does not allow harnessing the desired gain from decreasing $\rm{N}$ due to the low SE. On the bright side, however, the decreased SE can be mitigated by using the appropriate IM schemes instead of TIN. NOMA MAC/BC schemes achieve exactly this goal by using the IT-optimal schemes within each sub-topology, thus improving efficiency. This highlights the importance of using IM schemes for enabling a more efficient utilization of CN resources. To this end, we look at the problem from the following perspective that motivates this paper:

Current works aim at increasing the capacity of the CN by dividing it into P2P, MAC, or BC sub-topologies. That is, after $>$40 years of their discovery in IT, optimal schemes for MAC/BC have found their way to CNs as NOMA schemes. However, limiting our attention to MAC/BC schemes is not enough to support dense CNs, especially knowing that efficient IT-motivated IM schemes exist for a wider range of sub-topologies. From this point of view, it is wise to reap the benefits of such schemes sooner than later by involving sub-topologies for which mature IM schemes exist, such as interference channels (ICs), multi-way relay (MWR) channels, etc. By combining these sub-topologies with P2P/MAC/BC, the above conflict can be mitigated, which enables serving a larger number of UEs than what can be served using P2P/MAC/BC schemes using the same number of RBs, $\rm{N}$.

Given a dense CN, we would like to decompose CN into sub-topologies according to potential IM schemes. Hence, the following questions form our vision's skeleton:
\begin{tcolorbox}[colback=mavi!5!white,colframe=mavi!75!black]
\begin{enumerate}
\item[\textbf{Q1:}] \textbf{What rules should be used to compare sub-topologies?}
\item[\textbf{Q2:}] \textbf{How to decompose the CN into sub-topologies and allocate resources? }
\item[\textbf{Q3:}] \textbf{How to overcome potential practical challenges during the implementation?}
\end{enumerate}
\end{tcolorbox}
We call this \textit{`Network Topology Optimization'} since the CN is optimized by assigning UEs to different sub-topologies and allocating network resources to clusters within the sub-topologies. By judiciously hybridizing the master topology, the overall CN performance can be improved to a great extent. This is indeed the right time for this approach especially since future CNs will be ultra-dense and interference-limited, which is the suitable playground for potential IM schemes. 

\section*{ \centering\textbf{Network IT Approaches on Potential IM~Schemes and Sub-Topologies}}
IM schemes can handle interference using appropriate signal processing at the transceivers, other than avoiding it (OMA) or ignoring it (TIN). They reduce the impact of interference by allowing $\rm{N}$ to decrease while mitigating SE degradation, which in turn increases the overall CN capacity. Studying interference has a long history in IT, and was triggered by Shannon's earliest work on multi-terminal IT in 1961. This pioneering work led to an interest in multi-terminal networks and interference (e.g., MAC, BC, IC) by using superposition coding, joint decoding, or successive cancellation decoding (SCD) \cite{ElgamalKim}, which have already been considered in LTE-A. 

\begin{figure}
\includegraphics[width=\columnwidth]{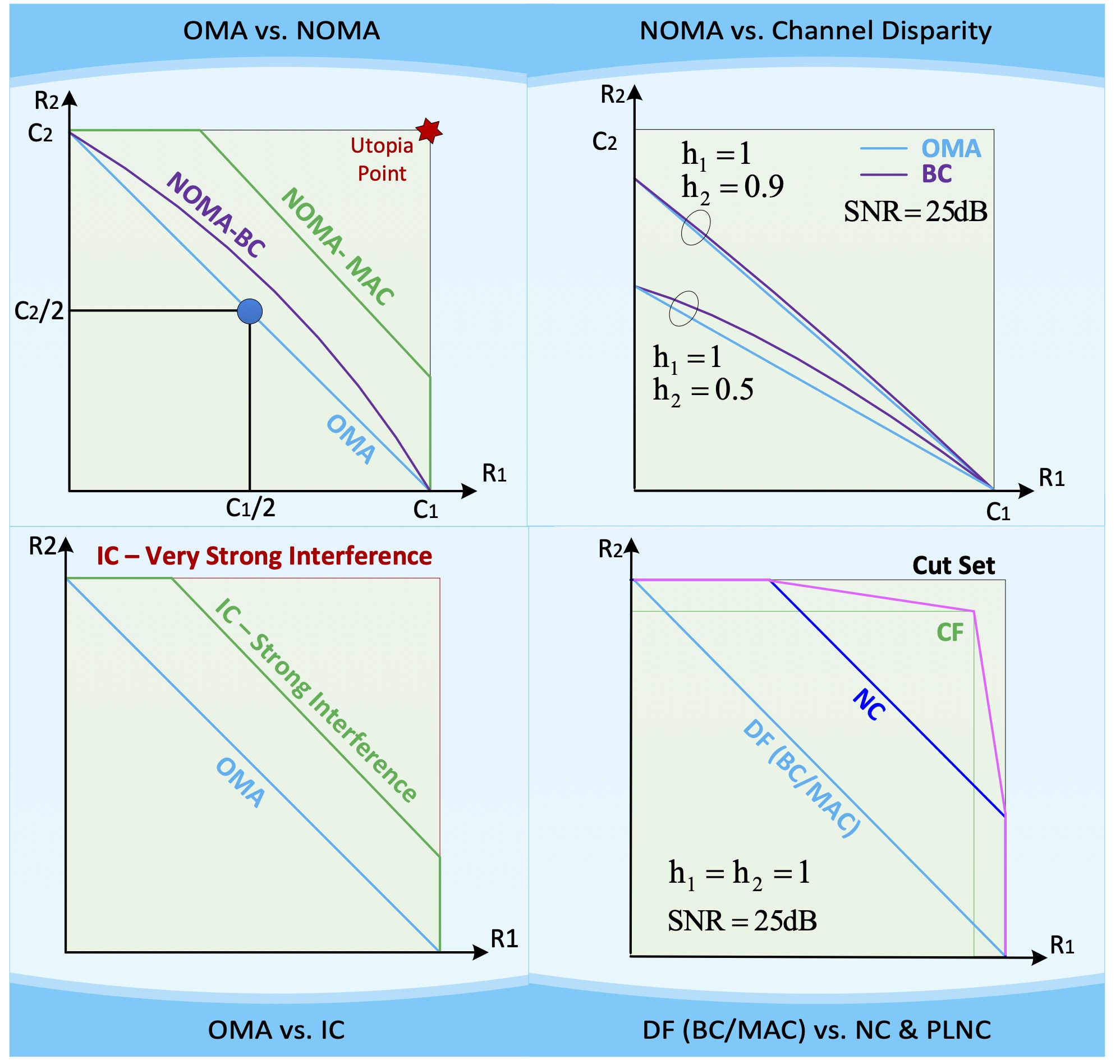}
\caption{Qualitative illustration of Gaussian channel achievable rate regions under different sub-topologies : OMA vs. NOMA \cite{ElgamalKim}, OMA vs. IC \cite{EtkinTseWang}, and DF vs. CF \& PLNC \cite{ChaabanSezgin_FnT}}
\label{fig:regions}
\end{figure}

The ultimate goal in a multiuser network is approaching the \textit{`Utopia Point'} in Fig. \ref{fig:regions}, which marks the achievable rate when interference is completely eliminated. While this might not be possible in general, it is possible in some cases thanks to IM. For instance, cell-edge UEs experiencing strong interference can benefit from IC schemes to a great extent (cf. Fig. \ref{fig:regions} left). In this case, IM enlarges the rate region from the triangular blue OMA region towards the utopia point, depending on the strength of interference. 

\begin{figure}
\includegraphics[width=\columnwidth]{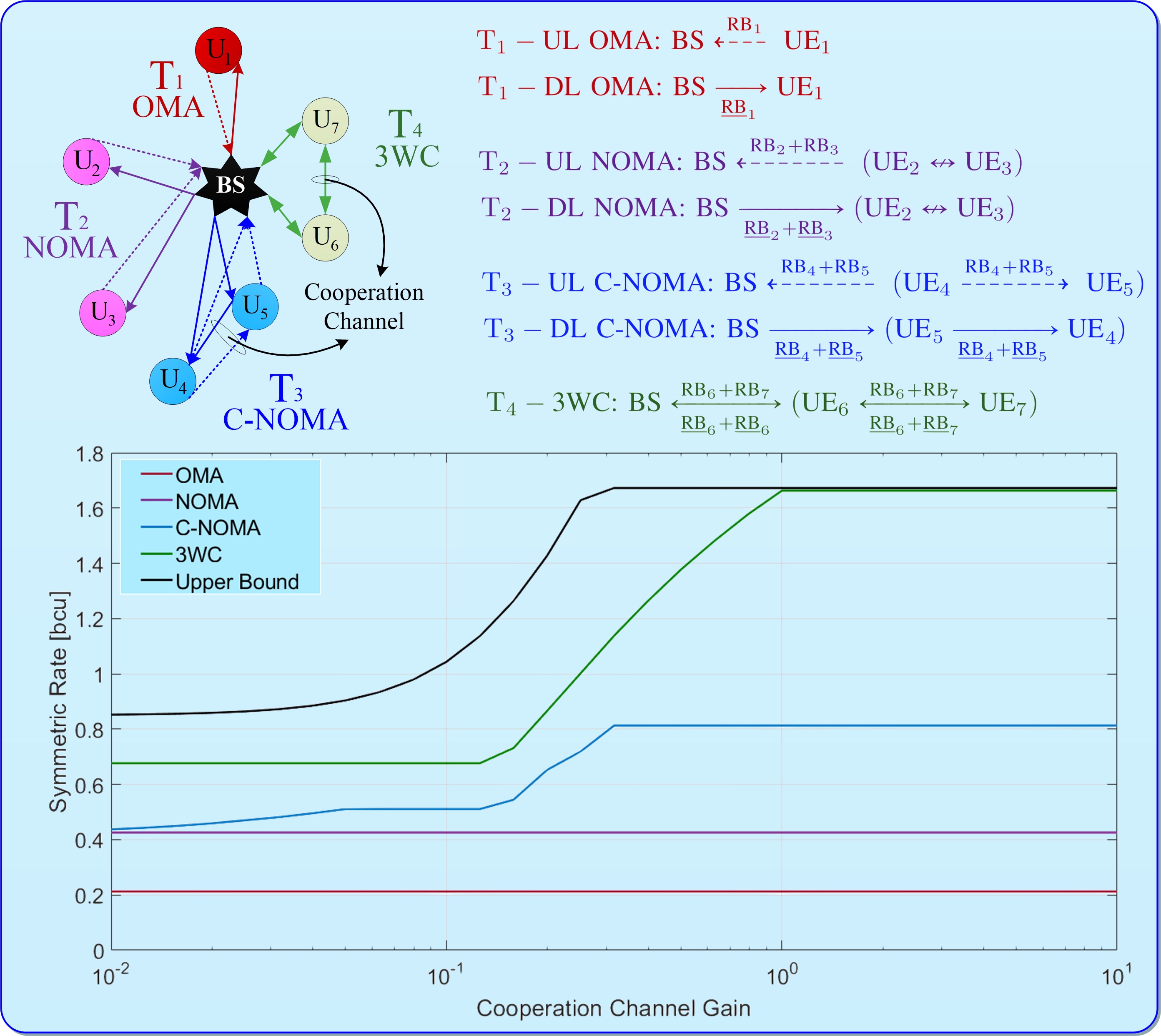}
\caption{Network partitioning demonstration and performance comparison of sub-topologies OMA, NOMA, cooperative NOMA (C-NOMA), and three-way channel (3WC).}  
\label{fig:plnc}
\end{figure}

Current NOMA solutions fall short of this aim due to a limitation that is best explained by the following toy example. Let us consider two UEs communicating with each other via a BS, where the network is operated as a BC and MAC in the DL and UL, respectively. The achievable rate region of the overall communication is the OMA triangle in Fig. \ref{fig:regions} (left) if OMA is used, and the intersection of the NOMA-BC and NOMA-MAC regions if NOMA is used. 
The slight improvement offered by NOMA will even vanish as the DL channel gains of two UEs gets closer. Eventually, the NOMA-BC and OMA regions will be equal if the channel gains are equalized. 
Even if the channel gains are different, a considerable gain can only be achieved at high signal-to-noise-ratio (SNR) if channel disparity is large. This scenario is highly likely in dense CNs where UEs have high SNR and comparable channels due to the cell size. Thus, only a small gain can be achieved using NOMA-BC schemes in such cases. A similar discussion applies to UL OMA and NOMA-MAC schemes. Although NOMA schemes decrease the required number of RBs per cell, they do not provide enough gain to support future dense CNs in this example. Note that the limitation here is not because of MAC/BC schemes since these are optimal for their respective channels; it is rather due to choosing to operate the whole network as a MAC/BC pair. 

Now, let us take a step back and ask the following question: \textit{Are BCs/MACs the only topologies with IM schemes that outperform OMA?} A quick review of network IT suggests the contrary. There are topologies with similar properties which are not exploited in state-of-the-art CNs such as two-hop relaying (THR) such as BS$\rightarrow$UE$_1$$\rightarrow$UE$_2$ or UE$_2$$\rightarrow$UE$_1$$\rightarrow$BS, or two-way relaying (TWR) such as BS$\leftrightarrow$UE$_1$$\leftrightarrow$UE$_2$. Under ideal conditions, the latter can exploit compute-forward (CF) relaying with lattice-codes, which can effectively double the achieved data rate, or even quadruple it if full-duplex (FD) transmission is allowed.

For instance, instead of a MAC/BC pair in the previous example, if we operate the network as TWR channel employing CF and physical-layer network-coding (PLNC) \cite{NamChungLee_IT}, we achieve rates which approach the Utopia point as shown in Fig. \ref{fig:regions} (right). This way, a significant gain can be achieved even if the channels are comparable. Interestingly, while NOMA is weaker when UEs have similar channels to the BS, PLNC-TWR is not, which enables achieving gains in cases where NOMA cannot. 

This combination also applies on a larger scale with more UEs as multi-way relaying (MWR) \cite{ChaabanSezgin_FnT}. MWR can also be useful for connecting small-cell BSs communicating with each other via a Macro-BS. In this case, CF relaying at the Macro-BS can bring significant gains compared to (N)OMA. Alternatively, two neighboring BSs can allow a set of UEs to share an RB in the UL (or DL), thus decreasing the total number of required RBs at the cost of increasing inter-cell interference. The result is a set of interfering MACs (or BCs) where IM by rate-splitting, decoding interference, or interference alignment (IA) with lattice codes can yield a far superior performance compared to the TIN. For an additional performance boost, alternating relaying can be used to enable FD communication, thus enabling a variety of FD topologies. Extra gains can be achieved with caching, that opens the doors for other IM schemes. But how to choose among these vast possibilities?

\section*{ \centering\textbf{Network Topology Optimization: \\ A Hybrid Multiple Access Perspective}} 
Current CNs do not allow accommodating this variety of schemes, thus wasting an opportunity to achieve superior performance. Therefore, it is crucial to obtain simple yet generic multiple access techniques suited for these schemes, in addition to measures and rules for partitioning nodes into sub-topologies of the desired structure. These points are further discussed below.

\begin{figure*}[t!]
\centering
\includegraphics[width=2 \columnwidth]{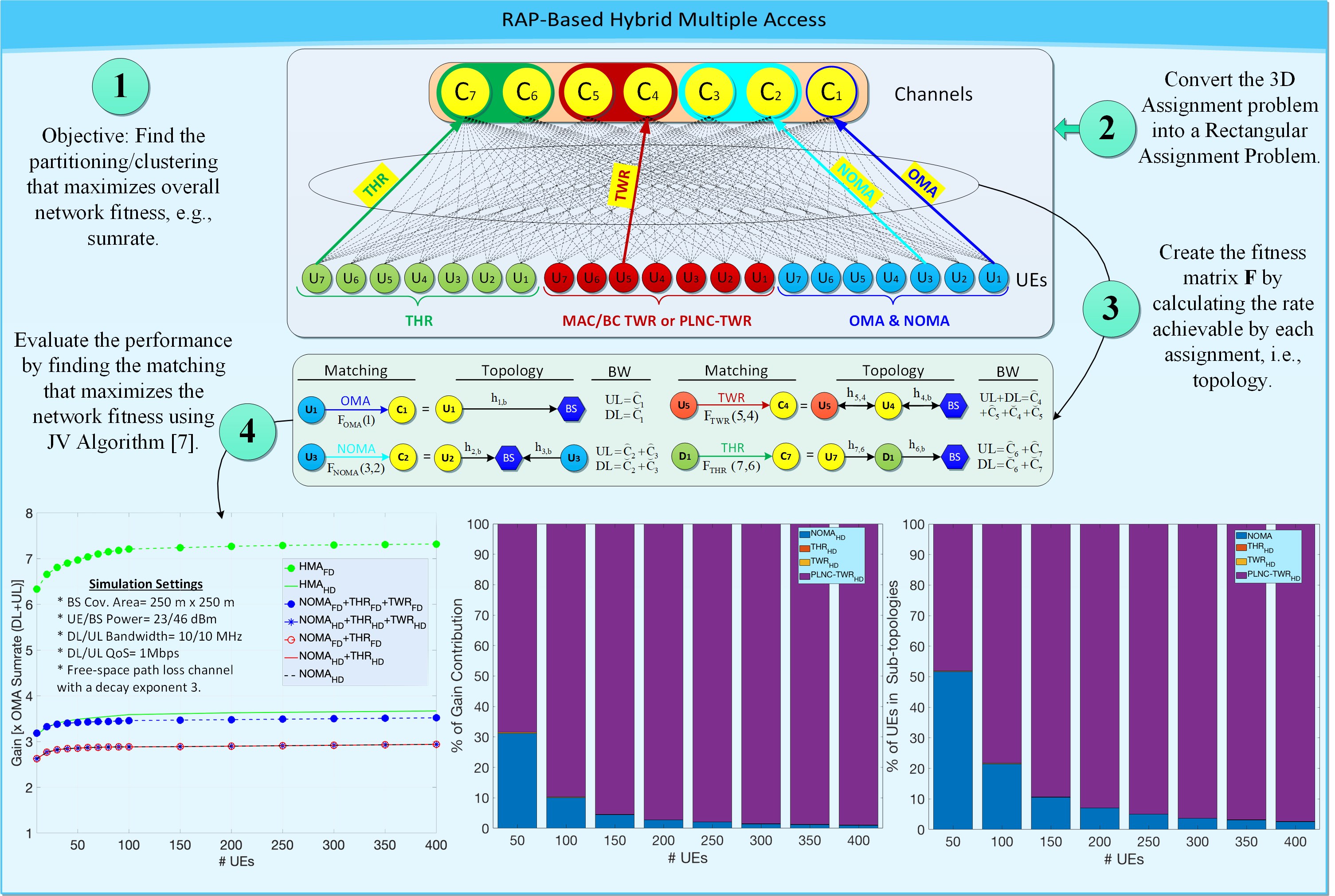}
\caption{Step-by-Step solution of HMA with $\sf{K}=\sf{M}=7$ and $\sf{N}=4$.}
\label{fig:bpm}
\end{figure*}

\subsection*{ \centering\textbf{Performance Measures and Rules of Thumb}}

Existing schemes in the IT literature are either customized to small networks which is insufficient for real-life CNs, or are highly sophisticated to be applied in a dense CN environment. Thus, it is necessary to scale-up the simplistic models and to scale-down the complexity of sophisticated IM schemes to be applicable in practical CNs. The challenge is to achieve these two goals while maintaining the schemes' spirit and gain. Once simplified, the schemes' performance can be characterized and compared in terms of the sum/symmetric rate or degrees-of-freedom. The next challenge is to design a framework for decomposing a CN into smaller sub-topologies. One could use the aforementioned performance criteria in topology optimization tools as a \textit{`fitness function'}. For example, if we decide to partition a CN as a set of sub-topologies $\{\sf T_i\}_{i=1}^N$ each with an associated fitness value $\sf F_i$, then the overall quality of this decision is a function of $\sf F_i$ such as $\sum_i\sf F_i$. By evaluating the fitness of each scheme/topology, a comparison will help to study the best way to break down a network into smaller sub-topologies. 

For a given sub-topology $\{\sf T_i\}$, $\sf F_i$ also depends on how its members clustered to share common network resources and how cluster resources are allocated among the cluster members. Therefore, $\sf F_i$ is jointly determined by the channel gains among the cluster members and optimization of cluster resources. If network resources are externally exploited by various clusters of different sub-topologies, the fitness of these coupled sub-topologies are determined by channel gains and resource allocation across these clusters. Thus, partitioning can be interpreted as selecting sub-matrices from the entire CN's channel gain matrix $\mathbf{M}$. Depending on the amount of interference represented by $\mathbf{M}$, one can decide whether to allow interference by using IM schemes or avoid interference using OMA. By using the aforementioned tailored IM schemes, it would be quite beneficial to develop rules of thumb that give strong hints on how to select these sub-matrices.

For $\sf N=4$, Fig. \ref{fig:plnc} illustrates partitions of a BS with 7 UEs, where UE$_i$, $i=1,\ldots,7$, is allocated with RB$_i$ and \underline{RB}$_i$ for UL and DL transmissions, respectively. A UE can individually exploit its dedicated RBs (i.e., OMA) or share it with another UE by operating on other topologies, e.g., NOMA, Cooperative-NOMA (C-NOMA), or three-way channel (3WC). For a generic couple of UEs, Fig. \ref{fig:plnc} also shows the fitness (UL+DL symmetric rates) of different topology types. Obviously, fitness increases as the cooperation channel gain gets stronger, and 3WC delivers a better performance than C-NOMA by utilizing the RBs more flexibly and efficiently.

\begin{table*}[htbp!]
\caption{Obtained percentage gains by various scheme combinations [OMA= \%100]}
\label{tab:gain}
\resizebox{\textwidth}{!}{%
\begin{tabular}{|c|l|l|l|l|l|l|l|l|l|c|c|c|c|c|c|c|c|}
\hline
\rowcolor[HTML]{96FFFB} 
\multicolumn{2}{|c|}{\cellcolor[HTML]{96FFFB}} &
  \multicolumn{16}{c|}{\cellcolor[HTML]{96FFFB}\textbf{\# UEs}} \\ \cline{3-18} 
\rowcolor[HTML]{96FFFB} 
\multicolumn{2}{|c|}{\multirow{-2}{*}{\cellcolor[HTML]{96FFFB}\textbf{Access Schemes}}} &
  \textbf{10} &
  \textbf{20} &
  \textbf{30} &
  \textbf{40} &
  \textbf{50} &
  \textbf{60} &
  \textbf{70} &
  \textbf{80} &
  \textbf{90} &
  \textbf{100} &
  \textbf{150} &
  \textbf{200} &
  \textbf{250} &
  \textbf{300} &
  \textbf{350} &
  \textbf{400} \\ \hline
\rowcolor[HTML]{D9F3FE} 
\cellcolor[HTML]{96FFFB} &
  \cellcolor[HTML]{96FFFB}{\color[HTML]{333333} \textbf{NOMA}} &
  262 &
  276 &
  281 &
  282 &
  283 &
  284 &
  285 &
  286 &
  287 &
  288 &
  289 &
  290 &
  291 &
  292 &
  293 &
  294 \\ \cline{2-18} 
\rowcolor[HTML]{D9F3FE} 
\cellcolor[HTML]{96FFFB} &
  \cellcolor[HTML]{96FFFB}{\color[HTML]{CB0000} \textbf{NOMA+THR}} &
  {\color[HTML]{CB0000} 262} &
  {\color[HTML]{CB0000} 276} &
  {\color[HTML]{CB0000} 281} &
  {\color[HTML]{CB0000} 282} &
  {\color[HTML]{CB0000} 283} &
  {\color[HTML]{CB0000} 284} &
  {\color[HTML]{CB0000} 285} &
  {\color[HTML]{CB0000} 286} &
  {\color[HTML]{CB0000} 287} &
  {\color[HTML]{CB0000} 288} &
  {\color[HTML]{CB0000} 289} &
  {\color[HTML]{CB0000} 290} &
  {\color[HTML]{CB0000} 291} &
  {\color[HTML]{CB0000} 292} &
  {\color[HTML]{CB0000} 293} &
  {\color[HTML]{CB0000} 294} \\ \cline{2-18} 
\rowcolor[HTML]{D9F3FE} 
\cellcolor[HTML]{96FFFB} &
  \cellcolor[HTML]{96FFFB}{\color[HTML]{3531FF} \textbf{NOMA+THR+TWR}} &
  {\color[HTML]{3531FF} 262} &
  {\color[HTML]{3531FF} 276} &
  {\color[HTML]{3531FF} 281} &
  {\color[HTML]{3531FF} 282} &
  {\color[HTML]{3531FF} 283} &
  {\color[HTML]{3531FF} 284} &
  {\color[HTML]{3531FF} 285} &
  {\color[HTML]{3531FF} 286} &
  {\color[HTML]{3531FF} 287} &
  {\color[HTML]{3531FF} 288} &
  {\color[HTML]{3531FF} 289} &
  {\color[HTML]{3531FF} 290} &
  {\color[HTML]{3531FF} 291} &
  {\color[HTML]{3531FF} 292} &
  {\color[HTML]{3531FF} 293} &
  {\color[HTML]{3531FF} 294} \\ \cline{2-18} 
\rowcolor[HTML]{D9F3FE} 
\multirow{-4}{*}{\cellcolor[HTML]{96FFFB}\rotatebox[origin=c]{90}{\textbf{HD}}} &
  \cellcolor[HTML]{96FFFB}{\color[HTML]{009901} \textbf{HMA}} &
  {\color[HTML]{009901} 315} &
  {\color[HTML]{009901} 331} &
  {\color[HTML]{009901} 340} &
  {\color[HTML]{009901} 345} &
  {\color[HTML]{009901} 348} &
  {\color[HTML]{009901} 351} &
  {\color[HTML]{009901} 353} &
  {\color[HTML]{009901} 355} &
  {\color[HTML]{009901} 357} &
  {\color[HTML]{009901} 359} &
  {\color[HTML]{009901} 361} &
  {\color[HTML]{009901} 363} &
  {\color[HTML]{009901} 364} &
  {\color[HTML]{009901} 365} &
  {\color[HTML]{009901} 366} &
  {\color[HTML]{009901} 367} \\ \hline
\rowcolor[HTML]{D9F3FE} 
\multicolumn{1}{|l|}{\cellcolor[HTML]{96FFFB}} &
  \cellcolor[HTML]{96FFFB}{\color[HTML]{CB0000} \textbf{NOMA+THR}} &
  {\color[HTML]{CB0000} 262} &
  {\color[HTML]{CB0000} 276} &
  {\color[HTML]{CB0000} 281} &
  {\color[HTML]{CB0000} 282} &
  {\color[HTML]{CB0000} 283} &
  {\color[HTML]{CB0000} 284} &
  {\color[HTML]{CB0000} 285} &
  {\color[HTML]{CB0000} 286} &
  {\color[HTML]{CB0000} 287} &
  {\color[HTML]{CB0000} 288} &
  {\color[HTML]{CB0000} 289} &
  {\color[HTML]{CB0000} 290} &
  {\color[HTML]{CB0000} 291} &
  {\color[HTML]{CB0000} 292} &
  {\color[HTML]{CB0000} 293} &
  {\color[HTML]{CB0000} 294} \\ \cline{2-18} 
\rowcolor[HTML]{D9F3FE} 
\multicolumn{1}{|l|}{\cellcolor[HTML]{96FFFB}} &
  \cellcolor[HTML]{96FFFB}{\color[HTML]{3531FF} \textbf{NOMA+THR+TWR}} &
  {\color[HTML]{3531FF} 318} &
  {\color[HTML]{3531FF} 332} &
  {\color[HTML]{3531FF} 338} &
  {\color[HTML]{3531FF} 340} &
  {\color[HTML]{3531FF} 342} &
  {\color[HTML]{3531FF} 343} &
  {\color[HTML]{3531FF} 344} &
  {\color[HTML]{3531FF} 345} &
  {\color[HTML]{3531FF} 346} &
  {\color[HTML]{3531FF} 347} &
  {\color[HTML]{3531FF} 348} &
  {\color[HTML]{3531FF} 349} &
  {\color[HTML]{3531FF} 350} &
  {\color[HTML]{3531FF} 351} &
  {\color[HTML]{3531FF} 352} &
  {\color[HTML]{3531FF} 353} \\ \cline{2-18} 
\rowcolor[HTML]{D9F3FE} 
\multicolumn{1}{|l|}{\multirow{-3}{*}{\cellcolor[HTML]{96FFFB}\rotatebox[origin=c]{90}{\textbf{FD}}}} &
  \cellcolor[HTML]{96FFFB}{\color[HTML]{009901} \textbf{HMA}} &
  {\color[HTML]{009901} 633} &
  {\color[HTML]{009901} 665} &
  {\color[HTML]{009901} 681} &
  {\color[HTML]{009901} 690} &
  {\color[HTML]{009901} 697} &
  {\color[HTML]{009901} 704} &
  {\color[HTML]{009901} 710} &
  {\color[HTML]{009901} 715} &
  {\color[HTML]{009901} 718} &
  {\color[HTML]{009901} 721} &
  {\color[HTML]{009901} 724} &
  {\color[HTML]{009901} 727} &
  {\color[HTML]{009901} 729} &
  {\color[HTML]{009901} 730} &
  {\color[HTML]{009901} 731} &
  {\color[HTML]{009901} 732} \\ \hline
\end{tabular}%
}
\end{table*}

\subsection*{ \centering\textbf{Hybrid Multiple Access}}
To accommodate different IM schemes, an HMA scheme is necessary to partition the network into sub-topologies which operate on different schemes. However, designing an optimal HMA falls within the class of mixed-integer non-linear programming (MINLP) problems which have an impractical time complexity (i.e., NP-hard) even for moderate-size CNs. To provide a deeper insight into the HMA design, we outline it as joint sub-problems of partitioning, clustering, and resource allocation.

\textit{Partitioning} aims to group UEs by sub-topology, i.e., each group includes UEs that 'favor' a specific topology and its corresponding IM scheme. Fig. \ref{fig:plnc} shows an example with four partitions. Generally, forcing the entire network to operate on NOMA cannot always be the best option since its achievable gain is limited by UEs' SCD efficiency \cite{Celik2019DistributedUL}. Likewise, NOMA may not be a good choice if you consider an objective other than the sum-rate, e.g., energy-efficiency or max-min rate. It is also possible to encounter problematic HMA scenarios where a cell-edge UE can neither achieve the desired rate by OMA nor can it be admitted to a NOMA cluster since it disadvantages existing cluster members. One way to accommodate such UEs is to exploit device-to-device communication based relaying via power-spectrum trading (PST). That is, devices with good channels but limited bandwidth can act as supporting relays for a UE in return for sharing its RB. 
Thus, an HMA scheme should be designed to partition the CN based on practical considerations such as channel matrix, hardware capabilities, fairness, QoS demands, etc. Notice that allowing UEs to join different sub-topologies would enhance the performance at the cost of a higher design complexity. 

\textit{Clustering} further divides these sub-topologies into groups of UEs that share common network resources. Clustering should wisely decide on cluster size and members; for instance, the spectral efficiency of NOMA schemes improves with the cluster size at the cost of increased complexity, decoding latency, and power consumption \cite{Celik2019DistributedUL}. On the other hand, cluster members should carefully be selected as per the channel matrix, since achievable NOMA gain is tightly connected with the channel gain disparity of UEs. Similarly, the complexity of MWR prohibits involving a large number of UEs; this is in addition to the complexity of relay selection that may become crucial for maximizing the performance.

\textit{Resource allocation} takes care of allocating CN resources (time, bandwidth, power) across sub-topologies and clusters, while satisfying UEs' QoS demands. HMA performance depends on how efficiently network resources are utilized, which is the primary motivation of research on network optimization. Unfortunately, existing resource allocation methods typically focus on a specific multiple access scheme, which is not be enough to reap the full benefits of our approach of diversifying and hybridizing potential IM schemes.  

To show the potential of HMA, we next probe an initial investigation on a centralized partitioning and clustering of $\sf M$ UEs within a macrocell. We assume that the number of channels $\sf K$ is equal to the number of UEs, the available bandwidth and BS power are equally distributed among the UEs, and channel $\left\{\mathsf{C_k} \right\}_{\sf k=1}^{\sf K}$, is initially dedicated to user $\mathsf{U_k}$ as in OMA. The problem is optimizing partitions/clusters to maximize the sum-rate of UEs, which can join at most one of four topologies ({$\sf N=4$}): OMA, NOMA, MAC/BC-TWR, and PLNC-TWR. Indeed, HMA is a three-dimensional assignment problem with a cost (fitness) matrix $\sf\bf{F}$ of size $\sf M \times M \times N$. As illustrated in Fig. \ref{fig:bpm}, we cast HMA as a rectangular assignment problem (RAP) by replicating UE vertices to address assignments between channels and different sub-topologies. A channel can be assigned at most one UE that implicitly determines both partition and cluster. The cost of assigning $\mathsf{C_k}$ to $\mathsf{U_k}$ is determined by the fitness function of the underlying topology, where fitness values are achievable rates after optimizing power and time-slot durations. Therefore, the optimal assignment will provide a desirable partitioning and clustering that optimize the overall fitness of the network. 

Fig. \ref{fig:bpm} shows the performance evaluation of the RAP-Based HMA scheme which is solved in polynomial time using the Jonker-Volgenant (JV) algorithm \cite{Jonker1987}. In bottom-left figure and its tabulated version in Table \ref{tab:gain}, HMA refers to the case that includes NOMA, THR, TWR, and PLNC-TWR. NOMA gain degrades as $\sf M$ increases because a higher user density affects the channel gain disparity within NOMA clusters. On the contrary, HMA schemes constantly improves as $\sf M$ increments since a dense network provides desirable relaying alternatives with better conditions. These can be seen from the bar-plots in Fig. \ref{fig:bpm} where the percentage of UEs operating on NOMA (PLNC-TWR) and their contribution to overall gain decreases (increases) with increasing densities. Notice that TWR is highly dominated by PLNC-TWR thanks to the high performance of the CF method. Finally, it is interesting to observe that HMA-FD delivers twice and five times more gain than HMA-HD and NOMA, respectively.

\section*{ \centering\textbf{Research Challenges and Opportunities}}

 
\subsection*{ \centering\textbf{Rethinking Network Virtualization}}

To realize the proposed HMA scheme, 6G networks should be able to orchestrate distinctive network functions of different sub-topologies combined into a single platform. Indeed, 5G already differs sharply from its predecessors in embracing software-defined reconfigurable architectures by means of two prominent technology trends: cloudification and virtualization. The cloudification has been explored in the realm of cloud radio access networks which interplay and integrate various aspects of signal processing, information theory, and networking. To accommodate distinctive QoS demands of different applications, network virtualization has been considered as a key enabler to divide cellular networks into slices dedicated for various vertical 5G industries. In a similar fashion, 6G networks should further consider set of UEs operating on the same sub-topology as a virtual sub-network and define specific virtual network functions that governs signal processing, communications, and networking components of the underlying access schemes.

\subsection*{ \centering\textbf{Potential IM Schemes}}
There are several schemes that have great promise when it comes to dense CNs, depending on the involved sub-topologies. Of particular interest are schemes which are cloud-enabled IM (CIM) schemes inspired from recent IC schemes in addition to relaying schemes.

 \textbf{CIM schemes} can be based upon IA and rate-splitting combined with common message decoding \cite{HuangCadambeJafar, EtkinTseWang}. Such schemes can provide significant gain but require BS coordination and feedback. Fortunately, coordination can be orchestrated by the cloud in cloud radio access networks (CRANs), with its high processing capabilities and global network view. The cloud can thus realize CIM schemes that handle clustering and IM jointly. In this context, it is insightful to note that a CRAN can be viewed as a BC in the infinite backhaul capacity extreme, and as an IC in the zero backhaul capacity extreme. This view calls for schemes that combine BC and IC schemes in the finite backhaul capacity regime. Earlier treatments mostly focused on the former (BC) with a combination of beamforming and TIN. Thus, there is a dire need to develop practical methods CIM beyond state-of-the-art techniques that use TIN \cite{TaoChenZhouYu}. This includes IA schemes motivated by cognitive radio networks \cite{HuangJafar_MIMOIC_Cooperation_Cognition}, X-networks \cite{HuangCadambeJafar}, coordinated multi-point transmission CoMP \cite{GamalAnnapureddyVeeravalli}, and CIM using rate-splitting and common message decoding \cite{Ahmad_etal}. Additionally, there is a need to investigate different cloud-enabled BS cooperation techniques (e.g., partial-message or compressed-signal sharing) in conjunction with the aforementioned CIM schemes while taking the cooperation cost into account.


 \textbf{Relaying schemes} of interest include THR, TWR and MWR, PLNC, FD communication, and noisy network-coding, all of which can bring considerable gains. Considering the vast relaying combinations in dense CNs, one can initially restrict attention to promising candidates such as sequential/parallel/one-hop relaying, and develop novel MWR schemes for these candidates. Fig. \ref{fig:bpm} shows an example of the great potential of PLNC-TWR in both its HD and FD forms.

Some of the IM schemes inspired by recent IT results may be impractical due to their complexity and sensitivity to channel-state information. Thus, it is important to simplify these IM schemes and to study the impact of finite-precision channel state information to bring these schemes closer to practice. To gain additional insight, it helps to derive IT upper bounds to gauge the performance of the simplified IM schemes and to discover conditions under which they provide considerable gains. Finally, these IM schemes should be fortified by clustering rules and algorithms for application in CNs under practical considerations such as cell size, transmit/receive power, number of antennas, and fading.

\subsection*{ \centering\textbf{Optimization Tools}}

\textbf{Matching Theory} provides mathematically tractable solutions for combinatorial problems of matching players in two partitions, e.g., UEs and channels in Fig. \ref{fig:bpm}. Depending on players' quota, matching problems can be classified as one-to-one matching, many-to-one matching, and many-to-many matching; which can be further categorized as matching with single or two-sided preferences. The deferred acceptance (DA) algorithm is a powerful matching procedure in which players iteratively make proposals which are either accepted or rejected by players of other partition respecting their preferences and quota \cite{Gale1962College}. Players make decisions based on individual information and preferences that can be represented by a matrix of fitness values whose columns and rows represent sub-topologies and users, respectively. These preferences can be customized as per UE's local information, objective, QoS demands, hardware capabilities, etc. Hence, DA is an inherently distributed and self-organizing algorithm that can tackle heterogeneity and complexity of CNs. 

However, \textit{externalities} of CNs pose extra challenges on matching problems, which can be interpreted as interdependencies among the players' preferences and decisions \cite{han2017matching}. Externalities can be interpreted as interference caused by other clusters operating at the same RBs. Hence, they are basically functions of the number of clusters and the identity of cluster members. These can be eliminated if UEs/clusters are assigned to dedicated resources, known as canonical matching. In one-to-one and one-to-many cases of canonical matching, DA always yields a stable matching \cite{Gale1962College}. Since the DA procedure does not necessarily yield a stable matching when externalities exist, it is important to extend the DA procedure to yield a stable matching by analyzing the channel matrix of users interested in a common resource. A proper design should also consider the resulting communication overhead due to the message passing among players and resources. 

\textbf{Game Theory} designs and analyzes the complex interactions among rational players. It is mainly classified as cooperative and non-cooperative based on collaborative and competitive inter-relations, respectively. Besides their superior performance, cooperative games are more suitable for network topology optimization as collaboration is inevitable for sophisticated IM schemes. There are two main types of cooperative games: coalition or network formation games. In this context, a coalition refers to a cluster whereas a partitions is a coalition of coalitions.


Coalition formation games are suitable for partitioning as they seek answers for optimal the coalition size and members by accounting for reduced gain due to the cooperation cost. 
Fitness functions and simplified rules of thumb can serve very well for coalition comparisons and merge/split decisions, respectively. If we set the OMA scheme as the initial set up, UEs can compare potential IM schemes, form partitions with UEs (merge) sharing similar characteristics, or leave (split) if they do not benefit from staying in a partition. Nonetheless, coalition formation games do not consider how the partition members are interconnected. 
Therefore, network formation games offer a more comprehensive approach for HMA since connectivity within the partition graphs determines interdependencies among the partitions. Network formation games can be based on myopic or far-sighted strategies to make decisions based on current network state or learning/predicting actions of other parties, respectively. 

\textbf{Auctioning Games} encompass bidders (buyers and/or sellers) and an auctioneer who collects bids to decide who will buy which items at what cost. Hence, auctioning is well-suited to PST-based relaying schemes where UEs with strong (weak) channels can bid for power (spectrum) of the UEs with weak (strong) channels. The auction starts with an initial price, which is iteratively updated by the auctioneer after collecting bids as long as supply exceeds the demand, and vice versa. Auctions are designed and categorized based on rules regulating bids, allocations, and payments \cite{auctioning}. In the HMA context, an allocation simply refers to a cluster where the buyers cooperate with sellers based on a payment (power or spectrum) agreement. In particular, combinatorial auctions are convenient for HMA schemes thanks to its generic form that allows bidders to place bids on several items. Hence, they can model complex clustering (relaying) scenarios where UEs can join multiple clusters as both spectrum sellers and power buyers. Moreover, the auctioneer can set a reserve price to keep the bidders in the auction, e.g., a bidder should perform better than its individual act (OMA) or QoS demand.


\textbf{Machine Learning} (ML) techniques are capable of adapting to the time-varying wireless environment and taking human behavior into account. This is possible via edge/swarm intelligence that is acquired by observing the environment, learning from history, predicting future network states, and accordingly taking necessary actions to continuously improve the performance. Thus, ML is an enabler of self-organizing, self-optimizing, and self-healing HMA schemes for beyond 5G.  

Although traditional ML techniques were studied for wireless communication applications, deep neural networks (DNNs) have recently boomed thanks to their ability to deal with problems that are highly complex to model and/or analytically intractable. As the depth increases, DNNs can provide a better level of abstraction to capture usually non-linear and dynamically changing relations. Thus, DNNs can deal with physical layer components such as channel estimation, coding, modulation, equalization. They have also been used for predicting traffic load, mobility pattern, and content interest. Hence, DNNs can be a powerful tool for UEs to design and evaluate fitness functions. Likewise, UEs can adapt their physical layer parameters to environmental changes, which may eliminate the unnecessarily frequent partitioning and clustering. Moreover, they can be used to build matching preferences, to make merge/split decisions, and to determine bidding values for maximum performance.   

Alternatively, a holistic approach would be use ML for the entire HMA scheme. In this respect, deep reinforcement learning (DRL) is appropriate for making decisions by learning from past interactions with the environment. For instance, a DRL-based HMA scheme can be designed by relating states to topologies, actions to merge/split operations, and rewards to fitness values, which is obtained by tailored IM schemes. Training this DRL-based HMA scheme on well-defined channel matrices could provide desirable performance. This approach could be especially beneficial in CIMs thanks to the global network view and computational capabilities in CRANs. 

\section*{ \centering\textbf{Conclusions}}
As classical communication paradigms approach their fundamental limits, future CNs should be designed in a more liberal and updated manner. This necessitates preferring IM schemes that allow interference rather than ignoring or avoiding it. After 40 years of their discovery, the telecom industry finally considers the transition from OMA to NOMA scheme. Although this paradigm shift sparkled by NOMA is appreciated, it still remains conservative in terms of topology types allowed in CNs. To this end, this paper shared our vision towards an HMA scheme that embraces different topologies. We also discussed open research areas on advanced IM schemes along with mathematical tools for network topology optimization. 

\bibliographystyle{IEEEtran}
\bibliography{final.bib}

\vspace*{-3\baselineskip}

\begin{IEEEbiography}[{\includegraphics[width=1.1 in,height=1.25in]{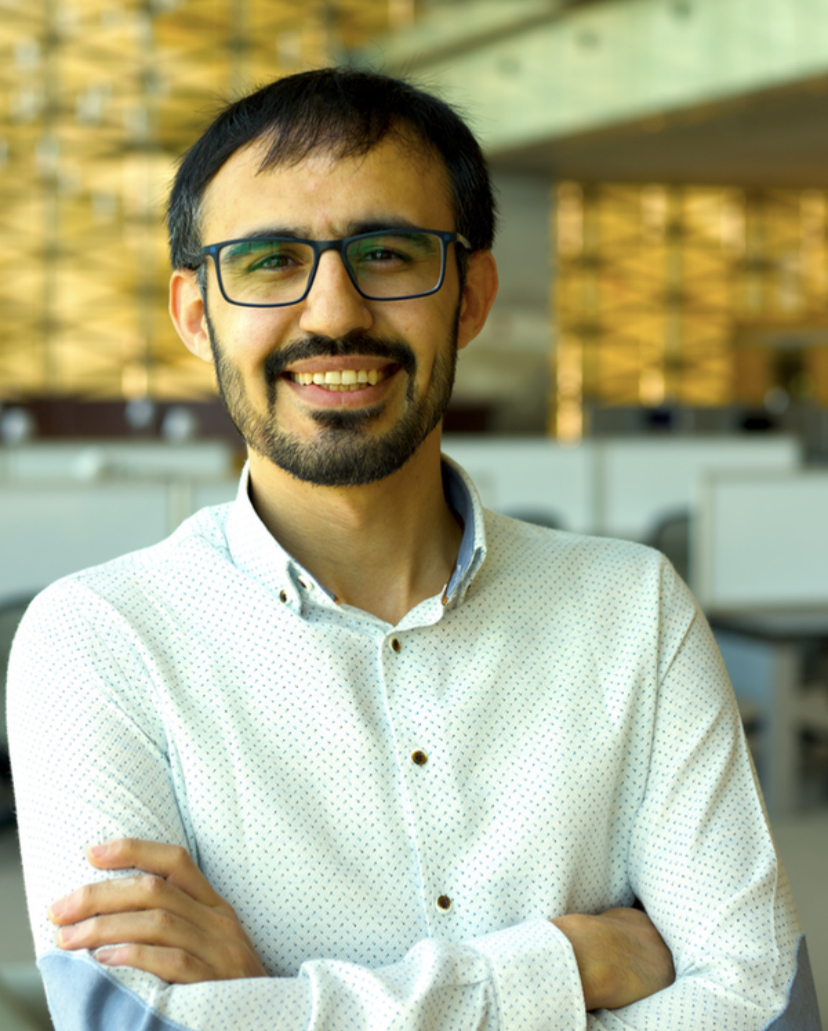}}]{Abdulkadir Celik}(S'14-M'16-SM'19) received the B.S. degree in electrical-electronics engineering from Selçuk University, Konya, Turkey, in 2009, the first M.S. degree in electrical engineering, the second M.S. degree in computer engineering, and the Ph.D. degree in co-majors of electrical engineering and computer engineering from Iowa State University, Ames, IA, USA, in 2013, 2015, and 2016, respectively. He was a post-doctoral fellow at King Abdullah University of Science and Technology (KAUST) from 2016 to 2020. He is currently a research scientist at communications and computing systems lab at KAUST. His research interests are in the areas of next-generation wireless communication systems and networks.
\end{IEEEbiography}

\vspace*{-5\baselineskip}
\begin{IEEEbiography}[{\includegraphics[width=1.1 in,height=1.25in]{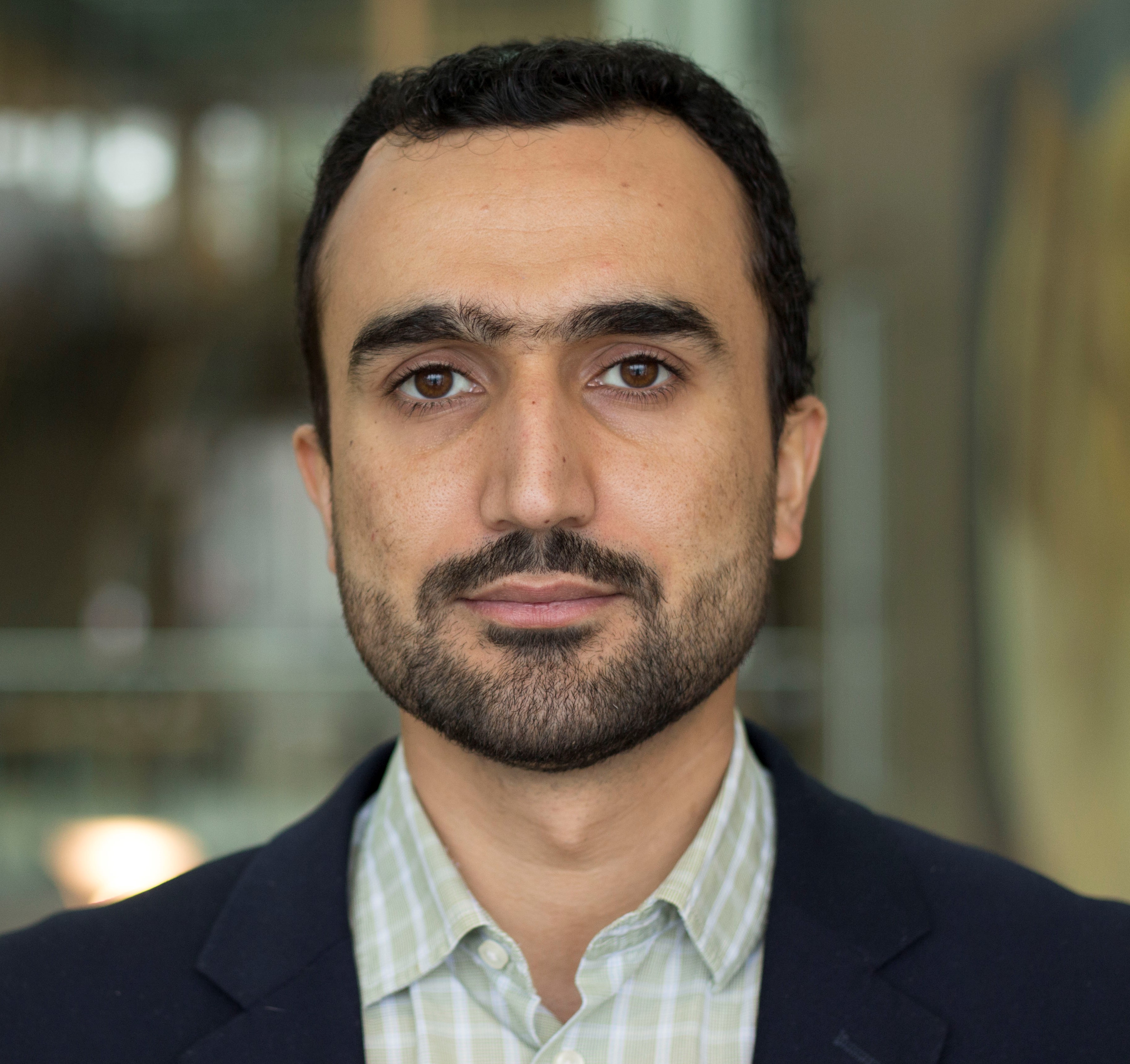}}]{Anas Chaaban}(S’09–M’14–SM’17) received the Ma{\^i}trise {\`e}s Sciences degree in electronics from Lebanese University, Lebanon, in 2006, the M.Sc. degree in communications technology and the Dr. Ing. (Ph.D.) degree in electrical engineering and information technology from the University of Ulm and the Ruhr-University of Bochum, Germany, in 2009 and 2013, respectively. From 2008 to 2009, he was with the Daimler AG Research Group On Machine Vision, Ulm, Germany. He was a Research Assistant with the Emmy-Noether Research Group on Wireless Networks, University of Ulm, Germany, from 2009 to 2011, which relocated to the Ruhr-University of Bochum in 2011. He was a Postdoctoral Researcher with the Ruhr-University of Bochum from 2013 to 2014, and with King Abdullah University of Science and Technology from 2015 to 2017. He joined the School of Engineering at the University of British Columbia as an Assistant Professor in 2018. His research interests are in the areas of information theory and wireless communications.
\end{IEEEbiography}

\vspace*{-5\baselineskip}
\begin{IEEEbiography}[{\includegraphics[width=1.1 in,height=1.25in]{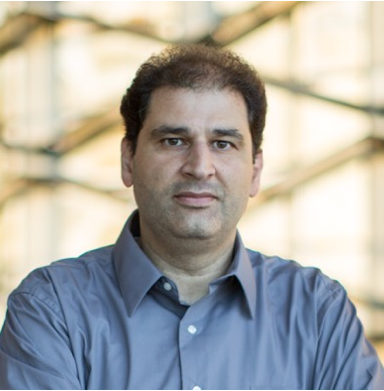}}]{Basem Shihada}(SM'12) is an associate 
\& founding professor in the Computer, Electrical and Mathematical Sciences \& Engineering (CEMSE) Division at King Abdullah University of Science and Technology (KAUST). He obtained his PhD in Computer Science from University of Waterloo. In 2009, he was appointed as visiting faculty in the Department of Computer Science, Stanford University. In 2012, he was elevated to the rank of Senior Member of IEEE. His current research covers a range of topics in energy and resource allocation in wired and wireless networks, software defined networking, internet of things, data networks, smart systems, network security, and cloud/fog computing.
\end{IEEEbiography}

\vspace*{-5\baselineskip}
\begin{IEEEbiography}[{\includegraphics[width=1.1 in,height=1.25in]{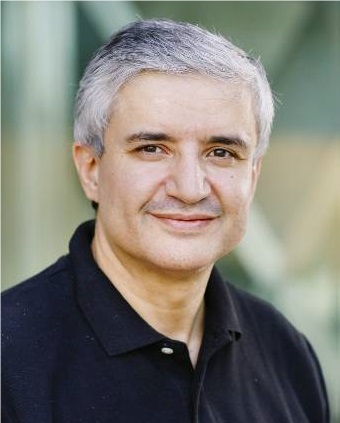}}]{Mohamed-Slim Alouini} (S'94-M'98-SM'03-F'09) born in Tunis, Tunisia. He received the Ph.D. degree in Electrical Engineering
from the California Institute of Technology (Caltech), Pasadena,
CA, USA, in 1998. He served as a faculty member in the University of Minnesota, Minneapolis, MN, USA, then in the Texas A\& M University at Qatar,
Education City, Doha, Qatar before joining King Abdullah University of
Science and Technology (KAUST), Thuwal, Makkah Province, Saudi
Arabia as a Professor of Electrical Engineering in 2009. His current
research interests include modeling, design, and
performance analysis of wireless communication systems.
\end{IEEEbiography}





\end{document}